\documentclass[reprint, amsmath,amssymb,aps,prl,showpacs]{revtex4-1}

\usepackage{graphicx}
\usepackage{dcolumn}
\usepackage{bm}

\begin{document}

\preprint{BAP'S/123-QED}

\title{Multiple-$q$ states and skyrmion lattice of the triangular-lattice Heisenberg antiferromagnet under magnetic fields}

\author{Tsuyoshi Okubo}
 \email{okubo@spin.ess.sci.osaka-u.ac.jp}
\author{Sungki Chung }%
\author{Hikaru Kawamura}%
\affiliation{Department of Earth and Space Science,
 Faculty of Science, Osaka University, Toyonaka, Osaka 560-0043, Japan%
}%
\date{\today}

\begin{abstract}
Ordering of the frustrated classical Heisenberg model on the triangular-lattice with an incommensurate spiral structure is studied under magnetic fields by means of a mean-field analysis and a Monte Carlo simulation. Several types of multiple-$q$ states including the {\it skyrmion-lattice} state is observed in addition to the standard single-$q$ state. In contrast to the Dzyaloshinskii-Moriya interaction driven system, the present model allows both skyrmions and anti-skyrmions, together with a new thermodynamic phase where skyrmion and anti-skyrmion lattices form a domain state.
\end{abstract}

\maketitle

Ordering of geometrically frustrated magnets has attracted recent interest \cite{Book1,Book2,Journal_topic}. The antiferromagnetic (AF) Heisenberg model on the triangular lattice is a typical example of two-dimensional (2D) geometrically frustrated magnets. When AF nearest-neighbor (NN) interaction only, the ground state of this system is the three-sublattice $120^\circ$ structure, which is {\it commensurate\/} to the underlying lattice. Under magnetic fields, the ordered state still keeps the three-sublattice structure and leads to a rich phase diagram \cite{Kawamura_Mag,Zhito_Mag}. 

When further-neighbor interactions become dominant, the ground state
often takes an {\it incommensurate\/} spiral structure. In recent experiments on the triangular-lattice AF $\mathrm{NiGa}_2\mathrm{S}_4$, an incommensurate spiral state due to strong further-neighbor interaction has been reported \cite{Nakatsuji,Nigas2,Nigas_rev}. In this compound,  magnitude of the AF third-neighbor interaction $|J_3|$ is larger than the ferromagnetic NN interaction $J_1$, $|J_3|/J_1 \sim  5$. In another compound $\mathrm{NiBr}_2$, an incommensurate state due to the AF third-neighbor and the ferromagnetic NN interactions were reported where $|J_3|/J_1 \sim 0.262$ \cite{NiBr2,NiBr2_inera}. 

From a symmetry viewpoint, an important difference of the incommensurate spiral from the $120^\circ$ structure is that the ground state possesses a three-fold degeneracy with respect to the choice of three equivalent directions of wavevectors on the lattice. Generally, a phase transition related to the breaking of such a discrete degeneracy could occur even in a 2D Heisenberg model. Indeed, the classical $J_1$-$J_3$ model in zero field exhibits a first-order transition associated with a breaking of such three-fold $C_3$ lattice symmetry. The ordered state is a single-$q$ state where one of three equivalent wavevector directions is chosen \cite{Tamura1,Tamura2, Balents}.  

 This three-fold degeneracy could also be a source of exotic ordered states, {\it e.g.\/}, various types of multiple-$q$ states where more than one wavevectors coexist. Although multiple-$q$ states were not reported in previous zero-field calculations, it might be realized under applied fields.  Multiple-$q$ states are generally incompatible with the fixed spin-length condition $|\bm{S}_i|=1$ and not favored at lower temperatures in the classical system, whereas they might be stabilized at moderate temperatures by thermal fluctuations. 

In this letter, based on a mean-field calculation and a Monte Carlo
simulation, we show that several types of multiple-$q$ states are indeed
stabilized under magnetic fields, and that one of them corresponds to
the so-called ``skyrmion-lattice'' state \cite{skyrmion_exp1,skyrmion_exp1_2,skyrmion_exp2,skyrmion_exp2_2,skyrmion_exp3,skyrmion_theory1,skyrmion_theory2,skyrmion_theory3,skyrmion_theory4,chiral_wave}. 

 We focus here on the triangular-lattice $J_1$-$J_3$ ior $J_1$-$J_2$ model jin a magnetic field of intensity $H$ whose Hamiltonian is given by  
\begin{equation}
 \mathcal{H}= -J_1\sum_{\left\langle i,j
\right\rangle}\bm{S}_i\cdot\bm{S}_j -J_{2,3}\sum_{\langle\langle i,j
\rangle\rangle}\bm{S}_i\cdot\bm{S}_j  -H\sum_i S_{i,z},
\label{Hamiltonian}
\end{equation}
where $\sum_{\langle i,j \rangle}$ and $\sum_{\langle\langle i,j
\rangle\rangle}$ mean the sum over the NN and the third-neighbor
(or  the second-neighbor) pairs, respectively. We consider classical
Heisenberg spins $\bm{S}_i=(S_{i,x}, S_{i,y}, S_{i,z})$ with
$|\bm{S}_i|=1$. An incommensurate ground state appears, in the
$J_1$-$J_3$ model, for the ferromagnetic NN interaction $J_1>0$ and the
AF third-neighbor interaction $J_3  <0$ with $ J_1/|J_3| < 4$, while, in
the $J_1$-$J_2$ model, for the AF $J_2$ with $-1< J_1/|J_2| < 3$.

 We first perform a mean-field analysis based on the method of Reimers {\it et al.} \cite{Reimers, Okubo}. Up to quartic order, the Landau free energy of the model is given by 
\begin{multline}
 \frac{F}{N} =
 \frac{1}{2}\sum_{\bm{q}}\left[3T-J_{\bm{q}}\right]\left|\bm{\Phi}_{\bm{q}}\right|^  2 -H\Phi_{\bm{0},z} \\
+  \frac{9T}{20}\sum_{\left\{\bm{q}\right\}}\hspace{-0.3em}\raisebox{0.4em}{$~^\prime$}\left[\bm{\Phi}_{\bm{q}_1}\cdot\bm{\Phi}_{\bm{q}_2}\right]\left[\bm{\Phi}_{\bm{q}_3}\cdot\bm{\Phi}_{\bm{q}_4}\right],
\end{multline}
where $\bm{\Phi}_{\bm{q}}$ is the order parameter corresponding to the  Fourier magnetization given by $ \bm{\Phi}_{\bm{q}} = \langle  \bm{S}_{\bm{q}} \rangle $ with  $\bm{S}_{\bm{q}} =  \frac{1}{N}\sum_{i}\bm{S}_{i}  \exp(-i\bm{q}\cdot\bm{r}_i)$ ($N$ the number of spins). The sum  $\sum_{\left\{\bm{q}\right\}}'$ runs over all $\bm{q}_i$s satisfying $\sum_{i}\bm{q}_i=\bm{0}$, and $J_{\bm{q}}$ is  the Fourier transform of the exchange interaction. 

 From the quadratic term of the free-energy expansion, one sees that the
 Fourier mode corresponding to the maximum $J_{\bm{q}}$ becomes unstable
 at $T_c = \frac{1}{3} J_{\bm{q}^\ast}$ where $\bm{q}^\ast$ is  the
 critical wavevector. In the $J_1$-$J_3$ model, $\bm{q}^\ast$  appears
 along the direction of NN bonds with
 $|\bm{q}^\ast|=\frac{2}{a}\cos^{-1}\left[\frac{1}{4}\left(1+\sqrt{1-\frac{2J_1}{J_3}}\right)\right]$
 ($a$ is the lattice constant),  while in the the $J_1$-$J_2$ model it
 appears along the second-neighbor direction  with
 $|\bm{q}^\ast|=\frac{2}{a\sqrt{3}}\cos^{-1}\left[-\frac{1}{2}\left(1+\frac{J_1}{J_2}\right)\right]$. Note
 that, reflecting the $C_3$ symmetry of the lattice, $\bm{q}^\ast$ is
 degenerate as $\pm \bm{q}_1^\ast$, $\pm \bm{q}_2^\ast$ and $\pm
 \bm{q}_3^\ast$.

 Just below the transition temperature, one might neglect all wavevectors other than the three critical modes $\bm{q}_i^\ast$ and the uniform $\bm{q}=0$ mode. Within this approximation, we find three metastable ordered states in addition to the paramagnetic state. These ordered states are characterized by the number of wavevectors appearing in the $xy$ component; they are single-$q$, double-$q$, and triple-$q$ states.

\noindent
(i) {\it The single-$q$ state\/}: Spins form an umbrella structure where the $xy$ component forms a single-$q$ spiral characterized by one of the three $\bm{q}_i^\ast$s, while the $z$ component consists of a uniform $q=0$ component $m_z$ along $H$.  This state is compatible with the condition $|\bm{S}_i|=1$. In fact, the ground state of our model turns out to be this single-$q$ state irrespective of the field intensity.

\noindent
(ii) {\it The double-$q$ state\/}: Here the $xy$ component is a superposition of two spirals with, {\it e.g,} $\pm \bm{q}_1^\ast$ and $\pm \bm{q}_2^\ast$, while the $z$ component forms a lineally polarized spin density wave with $\pm \bm{q}_3^\ast$ complementary to the $xy$ component with an additional uniform component.

\noindent
(iii) {\it The triple-$q$ state\/}: This state is a superposition of
three (distorted) spirals characterized by wavevectors $\bm{q}_1^\ast$,
$\bm{q}_2^\ast$, and $\bm{q}_3^\ast$ with an additional uniform component along $H$. In contrast to the single-$q$ and the double-$q$ states, three spiral planes are perpendicular to the $xy$ plane and rotate by $120$ degrees with each other. By using arbitrary three unit vectors lying on the $xy$ plane, $\bm{e}_i$ ($i=1,2,3$) satisfying $\sum_i\bm{e}_i=\bm{0}$, it is given by
\begin{subequations}
\begin{align}
 \bm{S}_{i,xy}& = I_{xy}\sum_{j=1}^3
 \sin(\bm{q}^\ast_j\cdot\bm{r}_i+\theta_j)\bm{e}_j, \label{skyrmion_lattice_xy}
\\
 S_{i,z} &= I_z\sum_{j=1}^3 \cos(\bm{q}^\ast_j\cdot\bm{r}_i+\theta_j) + m_z, 
\label{skyrmion_lattice_z}
\end{align}
\end{subequations}
where we introduced the phase factors $\theta_i$ ($i=1 \sim 3$)
satisfying the constraint $\cos(\theta_1+\theta_2+\theta_3)=-1$,
$I_{xy}$ and $I_z$ being $T$-dependent constants. 

 Interestingly, this triple-$q$ state spin configuration just
 corresponds to the ``{\it skyrmion lattice}'' recently discussed in
 conjunction with several ferromagnetic compounds MnSi, FeCoSi, and FeGe
 under magnetic fields
 \cite{skyrmion_exp1,skyrmion_exp1_2,skyrmion_exp2,skyrmion_exp2_2}. In
 these compounds, the skyrmion lattice is stabilized by the
 anti-symmetric Dzyaloshinskii-Moriya (DM) interaction. In contrast, the
 skyrmion lattice in the triple-$q$ state of the present model is
 realized via the frustrated symmetric exchange interaction. Note that
 skyrmions of our model can take both signs, {\it e.g.\/}, skyrmions and
 anti-skyrmions, because our Hamiltonian \eqref{Hamiltonian} keeps the
 mirror symmetry in the $xy$ spin component. Such $Z_2$ symmetry of the
 Hamiltonian, which is absent in the DM system, is spontaneously broken
 in the triple-$q$ state. Note that the direction of $\bm{e}_i$ in
 Eq. (3) is independent to the wavevector $\bm{q}_i^\ast$. In case of
 the DM system, on the contrary, $\bm{e}_i$ is fixed to be perpendicular to
 $\bm{q}_i^\ast$ to minimize the DM interaction, leading to a helix. 

 In the triple-$q$ state, the state keeps the $C_3$ symmetry of the lattice, while it is broken in the single-$q$ and the double-$q$ states. More precisely, $C_3$ symmetry is broken only in the $xy$ component in the single-$q$ state, whereas it is broken both in the $xy$ and $z$ components in the double-$q$ state. The change in $\theta_i$ of Eq.(3) induces a translation of the skyrmion lattice.

\begin{figure}
  \includegraphics[width=8cm]{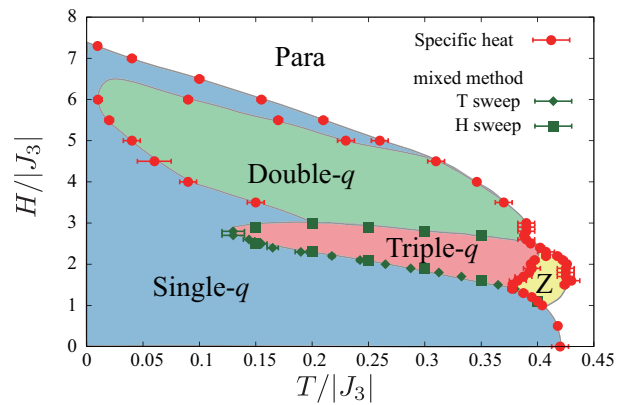}
 \caption{(color online).  Phase diagram of the $J_1$-$J_3$ model with $J_1/J_3 = -1/3$ in the temperature-field plane, obtained by a Monte Carlo simulation. Transition temperatures between the paramagnetic and the ordered phases are determined from the specific-heat-peak position, while those between the triple-$q$ and the single-$q$ (or the double-$q$) phases are determined by the mixed-phase method \cite{mixed_phase1}.}
 \label{fig_phase}
\end{figure}

 By comparing the free-energy of these ordered states, we construct a mean-field phase diagram in the $T$-$H$ plane. We find that the single-$q$ state always has a lower free energy than those of the double-$q$ and the triple-$q$ states. Thus, within the mean-field approximation, multiple-$q$ states are not stabilized. Meanwhile, we find that the free-energy difference between the single-$q$ and the triple-$q$ states becomes small at moderate fields, similarly to the case of a mean-field analysis of the DM ferromagnet \cite{skyrmion_exp1}. Hence, the fluctuation effect not taken into account in the mean-field approximation might eventually stabilize the multiple-$q$ states. 

 In order to further clarify the situation, we perform a Monte Carlo simulation based on the standard heat-bath method combined with the over-relaxation method. Our unit MC step consists of one heat-bath sweep and ten over-relaxation sweeps. The lattice is a $L\times L$ triangular lattice with $36 \le L \le 288$ with periodic boundary conditions. Typically, a single run contains $2 \sim 4 \times 10^5$ MC steps per spin at each temperature, while averages are made over $3\sim 5$ independent runs. 

 The resulting  $T$-$H$ phase diagram of the $J_1$-$J_3$ model is shown in Fig.~\ref{fig_phase} for $J_1/J_3=-1/3$. We find that, in addition to single-$q$ state, the double-$q$ and the triple-$q$ states are stabilized under magnetic fields due to fluctuations. In the low temperature limit, the single-$q$ state is always stable consistently with the the fixed spin-length condition. At larger fields, the single-$q$ phase cuts in between the paramagnetic and the double-$q$ phases. A similar phase diagram is obtained also for the $J_1$-$J_2$ model (see the online supplement \cite{supplement}).

\begin{figure}
  \includegraphics[width=8.5cm]{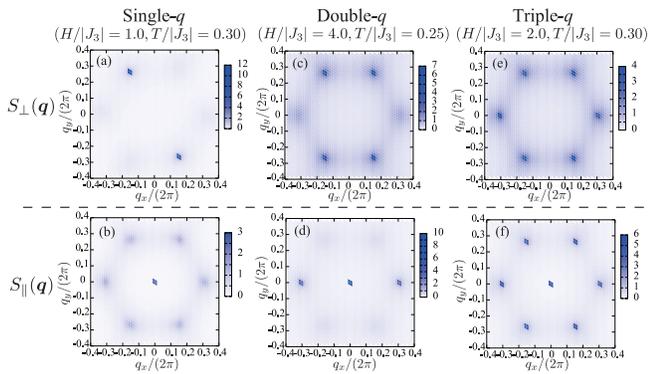}
 \caption{(color online). 
The intensity plot of the spin structure factor for $J_1/J_3 = -1/3$, in
 the single-$q$ phase (a, b), the double-$q$ phase (c, d), and the
 triple-$q$ phase (e, f). The lattice size is $L=72$. The upper and lower figures represent the spin structure factors of the $xy$ component $S_\perp(\bm{q}) =
\frac{1}{N} \sum_{\mu=x,y}\langle|\sum_i
 S_{i,\mu}e^{-i\bm{q}\cdot\bm{r}_i}|^2\rangle$ and of the $z$ component
 $S_\parallel(\bm{q}) = \frac{1}{N}\langle|\sum_i
 S_{i,z}e^{-i\bm{q}\cdot\bm{r}_i}|^2\rangle$. The gray (color) scale
 represents $\sqrt{S_{\perp, \parallel}(\bm{q})}$. }
  \label{fig_struct}
\end{figure}

The computed spin structure factors are shown in Fig.~\ref{fig_struct} for each case of the single-$q$, the double-$q$ and the triple-$q$ phases. As can be seen from the figure, all qualitatively features of the mean-field analysis are met here. It should be noticed that sharp spots observed at $\bm{q}=\bm{q}_i^\ast$ are not true Bragg peaks because we are dealing with the 2D Heisenberg model. If one recalls that the ordered states possess a continuous degeneracy due to the $U(1)$ symmetry associated with spin rotations (around $z$) and translations, the observed sharp spots should be quasi-Bragg spots associated with power-law spin correlations.

\begin{figure}
  \includegraphics[width=8.5cm]{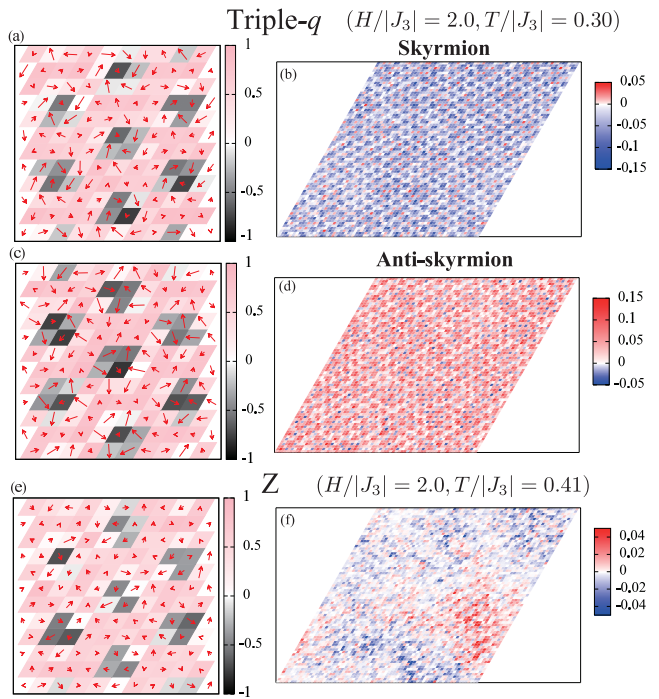}
 \caption{(color online). Typical real-space spin configurations (left figures) and the intensity plots of the skyrmion density (right figures), in the triple-$q$ phase (a-d) and in the $Z$ phase (e, f). Upper (middle) figures represent the skyrmion (anti-skyrmion) lattice. The $xy$ component of spins is represented by the arrow, while the $z$ component is given by the gray (color) scale. Short-time average over $10$ MCS is made to reduce the thermal noise. The lattice size is $L=72$.}
 \label{fig_snap}
\end{figure}

 We show in Fig.~\ref{fig_snap}(a) and (c) typical real-space spin
 configurations obtained in our MC simulation for the triple-$q$
 state. One can see that the $z$ component, shown by the gray (color)
 scale, often takes an opposite direction to the field direction forming
 a triangular superlattice expressed by balck color. The $xy$ component
 represented by arrows forms a vortex (a) or an anti-vortex (c) pattern
 around the black spots. Such a configuration is indeed the skyrmion (or
 anti-skyrmion) lattice predicted by the mean-field calculation as a
 metastable structure (3). In Fig.~\ref{fig_snap}(b) and (d), we show the corresponding skyrmion-density pattern. The local skyrmion density is defined here as the directed area of the sphere surface spanned by three spins on every elementary triangle on the lattice \cite{skyrmion_density}. As can be seen from the figure, in the triple-$q$ state the skyrmion (or anti-skyrmion) forms a triangular-lattice with a lattice spacing $\frac{4\pi}{\sqrt{3}|\bm{q}^\ast|}$. The total sum of the skyrmion density over the entire system is non-zero, being negative for the skyrmion lattice and positive for the anti-skyrmion lattice. In the single-$q$ and the double-$q$ phases, this sum turns out to vanish.

\begin{figure}
  \includegraphics[width=8.5cm]{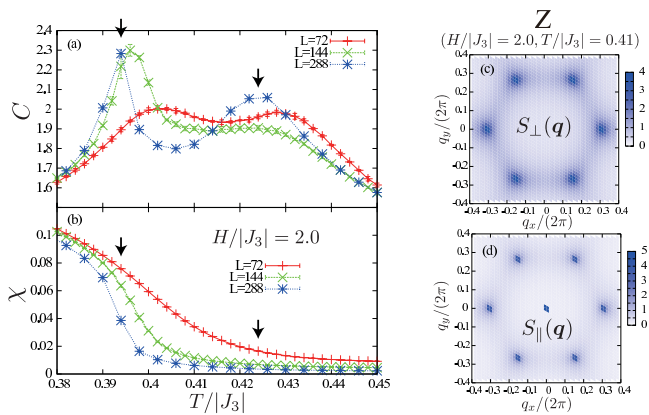}
 \caption{(color online). The temperature dependence of the specific heat per spin (a) and of the total scalar chirality per plaquette (b) for $H/|J_3|=2.0$. The spin structure factor of the $xy$ component (c) and of the $z$ component (d) in the $Z$ phase. $J_1/J_3=-1/3$ in all cases.}
 \label{fig_heat}
\end{figure}

 The $T$-$H$ phase diagram of Fig.~\ref{fig_phase} contains a new phase labeled $Z$,  right to the triple-$q$ phase, which is not predicted in the mean-field
 analysis even as a metastable state. Its existence is suggested from, {\it e.g.\/}, the specific-heat shown in Fig.~\ref{fig_heat}(a), where clear double peaks are observed at $\simeq 0.39 |J_3|$ and $\simeq 0.42 |J_3|$. The total scalar chirality, $\chi \equiv \sqrt {\langle (\frac{1}{2N}\sum_i \chi_i)^2\rangle}$ with $\chi_i=\bm{S}_{i_1}\cdot (\bm{S}_{i_2}\times \bm{S}_{i_3})$ where $i_1\sim i_3$ are three sites on an elementary triangle $i$ (both upward and downward), can be regarded as an order parameter of the $Z_2$ mirror symmetry. As shown in Fig.~\ref{fig_heat}(b), $\chi$ grows at a lower transition temperature, indicating that the  $Z_2$ mirror symmetry is preserved in the $Z$ phase.

 A typical spin configuration in the $Z$ phase is shown in Fig.~\ref{fig_snap}(e). The $z$ component forms a triangular superlattice similar to the triple-$q$ phase, whereas the $xy$ component remains disordered (paramagnetic). The corresponding spin structure factors shown in Fig.~\ref{fig_heat}(c) and (d) also indicate that the $xy$ component exhibits very broad spots in contrast to the sharp spots of the $z$ component. Thus, in the $Z$ phase only the $z$ component retains a quasi-long-range order similar to the triple-$q$ phase, while the $xy$ component remains disordered with a finite correlation length of about $20$ lattice spacings. Skyrmion density pattern of the $Z$ phase is shown in Fig.~\ref{fig_snap}(f), where one sees that skyrmions and anti-skyrmions are mixed in the form of domains. The domains size is comparable to the transverse ($xy$) spin correlation length deduced from the width of the peak of $S_\perp(\bm{q})$ (Fig.~\ref{fig_heat}(c)). Thus, the $Z$ phase is a domain state consisting of both skyrmion and anti-skyrmion lattices, where the total skyrmion number remains zero. At the para-$Z$ transition, the $U(1)\times U(1)$ symmetry associated with the two independent $\theta_i$ variables of Eq.(3) is broken in an algebraic manner. This observation suggests that the para-$Z$ transition is actually of the Kosterlitz-Thouless type.

 We now wish to compare the skyrmion-lattice state of the present model with that of the DM system as observed in MnSi, FeCoSi or FeGe \cite{skyrmion_exp1,skyrmion_exp1_2,skyrmion_exp2,skyrmion_exp2_2}. There are two important differences between the skyrmion lattice of the two systems: (i) First, the skyrmion lattice of the present model is much denser than its DM counterpart, with smaller lattice constant of $\sim 2\sqrt{3} a$ (see Fig.~\ref{fig_snap}). In the DM-driven system, the lattice constant is typically an order of magnitude larger, because the DM interaction is usually considerably weaker than the dominant exchange interaction.  Since the anomalous Hall conductivity due to skyrmion texture is proportional to the skyrmion density \cite{Nagaosa}, larger Hall conductivity is expected if the skyrmion (or anti-skyrmion) lattice in the triple-$q$ state could be stabilized in appropriate metallic materials by the present mechanism. (ii) Second, the $Z_2$ mirror symmetry is absent in the DM system, while it is kept as a Hamiltonian symmetry in the present model, being spontaneously broken in the triple-$q$ state. Its consequence is that both skyrmion and anti-skyrmion lattices, inter-connected via the $Z_2$ symmetry, are possible in the present model, together with a new $Z$ phase where skyrmion and anti-skyrmion lattices form a domain state. 

 Triangular-lattice compounds $\mathrm{NiGa}_2\mathrm{S}_4$ and  $\mathrm{NiBr}_2$ might be candidates of the multiple-$q$ states. Although  the observed zero-field properties of $\mathrm{NiGa}_2\mathrm{S}_4$ seem not consistent with a simple classical $J_1$-$J_3$ model in view of the absence of an expected first-order transition  \cite{Nigas_rev,Tamura1,Tamura2,Balents}, its local spin structure is  certainly an incommensurate spiral. Then, the multiple-$q$ structures might possibly be formed under magnetic fields, the required field roughly estimated to be $30\sim 50$ T. In case of $\mathrm{NiBr}_2$, estimated value $|J_3|/J_1 \simeq 0.262$ \cite{NiBr2_inera} is much less than that of the present study $|J_3|/J_1 =3$. Although larger $|J_3|/J_1$ seems favorable to the formation of the triple-$q$ state, further theoretical and experimental studies are desirable to clarify the general dependence on $|J_3|/J_1$.

 We finally note that within a mean-field approximation the multiple-$q$ states can be obtained as metastable states only by assuming the three-fold degeneracy of the ordered state. This suggests that the multiple-$q$ states and the skyrmion lattice could be realized not only in the triangular lattice, but also in other lattices with a trigonal symmetry, {\it e.g.}, the honeycomb and the kagome lattices. 

 In summary, we studied the ordering of triangular-lattice Heisenberg magnets with an incommensurate spiral structure. On the basis of a mean-field analysis and a Monte Carlo simulation, we found several multiple-$q$ phases under magnetic fields in addition to the standard single-$q$ phase. The spin structure in the triple-$q$ phase is the skyrmion (or anti-skyrmion) lattice. In contrast to the DM-induced skyrmion lattice, the present model keeps the $Z_2$ mirror symmetry, which enables both skyrmions/anti-skyrmions and gives rise to a new $Z$ phase, a domain state consisting of skyrmion and anti-skyrmion lattices.

\acknowledgments

The authors are thankful to T. Arima, Y. Tokura, S. Onoda, Y. Onose for
useful discussion. This study was supported by Grand-in-Aid for
Scientific Research on Priority Areas ``Novel States of Matter Induced
by Frustration''(19052006). We thank Supercomputer Center, ISSP,
University of Tokyo for providing us with the CPU time.

\begin{figure*}
  \includegraphics{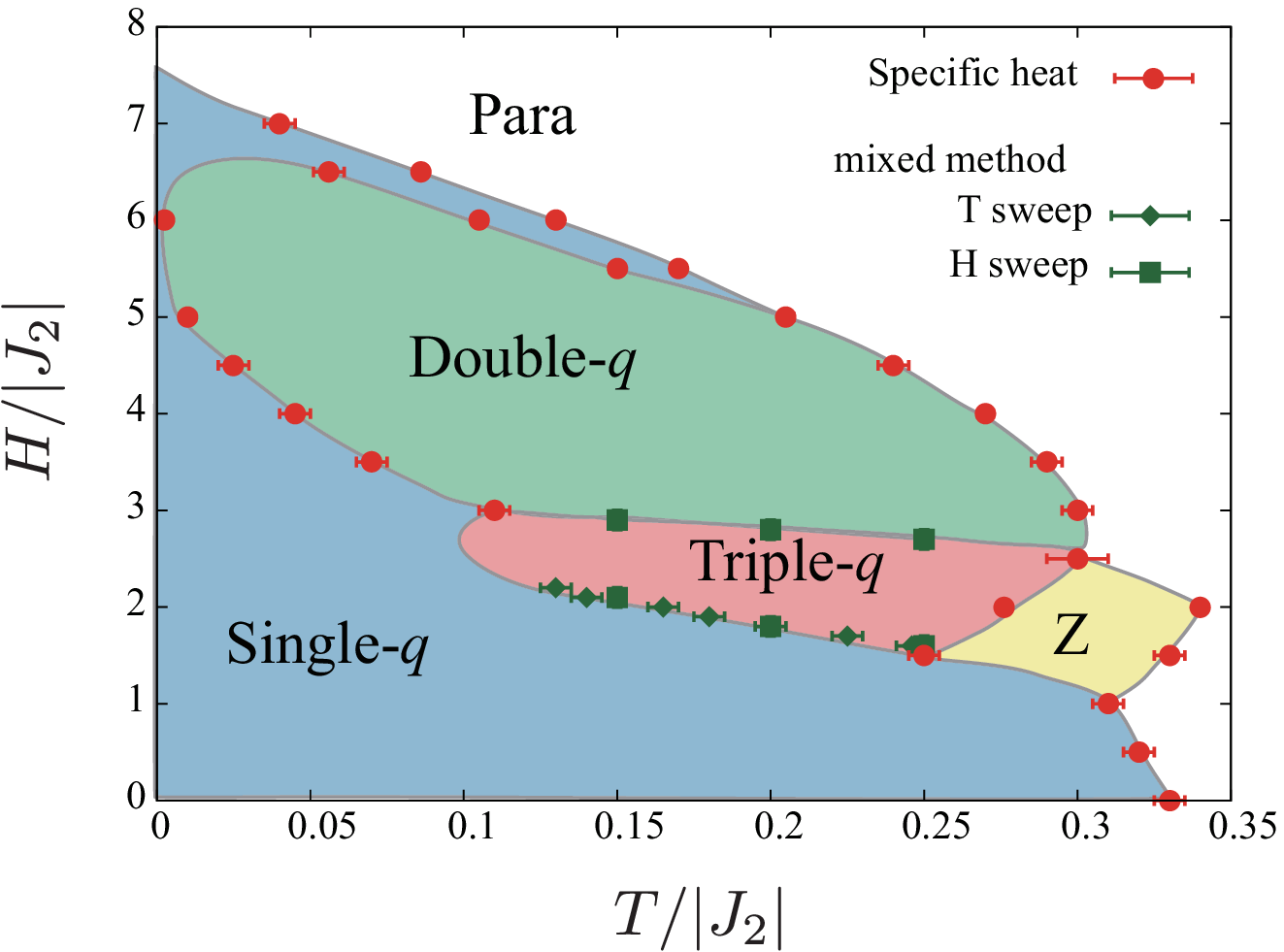}
 \caption{Phase diagram of the $J_1-J_2$ model with $J_1/J_2 = -1/4$ in
 the temperature-field plane, obtained by a Monte Carlo simulation. The
 Hamiltonian of the model is given by Eq.~(1) in the main text. Transition temperatures between the paramagnetic and the ordered phases are determined from the specific-heat-peak position, while those between the triple-$q$ and the single-$q$ (or the double-$q$) phases are determined by the mixed-phase method.}
\end{figure*}
\end{document}